-
\mag=1200
\vsize=9.3truein
\hsize = 6.3 truein
\parskip 2ex plus .5ex minus .1ex
\parindent=0.0truein
\headline{\hfil \folio \hfil}
\footline{\hfil \null}

\def \nl{\hfil\break}
\def \s{\sum_{n=1}^{\infty}}
\def \d{{\rm d}}
\def \f#1#2{{#1\over{#2}}}

\def \ip{\int_0^{\pi}}
\def \ip4{\int_0^{\pi/4}}
\def \ip2{\int_0^{\pi/2}}
\def \i1{\int_0^1}
\def \if{\int_0^{\infty}}

\def \p#1{\f{1}{\pi^#1}}

\def \rchi{\raise1.8pt\hbox{$\chi$}}
\def \ch#1{\rchi_{#1}}

\def \L{$L$-series }
\def \w{\omega}
\def \ww{\w^2}
\def \p{\phi}
\def \ll{(1-2^{-s})L_1(s)L_{-4}(s)}
\newcount\nexteq \nexteq=1
\newcount\sect \sect=1
\baselineskip=22 truept
\def\eqnn{(\the\sect.\the\nexteq)}

\def\eqn{\eqno{\eqnn}\global\advance\nexteq by 1} %generate and advance
\def\eq{{\eqnn}\global\advance\nexteq by 1} %for use in eqalignno
{\centerline{\bf Dirichlet L-series with real and complex characters}}
\centerline {\bf {and their application to solving double sums.}}

I.J.Zucker$^{1}$ and R.C.McPhedran$^2$.

{\it 1. Wheatstone Physics Laboratory, King's College, University of London, Strand, London WC2R 2LS, U.K. }
\nl {\it 2. School of Physics, University of Sydney, NSW 2006, Australia.}

\centerline {\bf {Abstract.}}

A description of the properties of \L with complex characters is given.  By using these, together with the more familiar \L with real characters, it is shown how certain two dimensional lattice sums, which previously could not be put into closed form, may now be expressed in this way.

\centerline {\bf{I.Introduction}}
Lattice sums are expressions of the form
$$
\sum_{\bf l}F({\bf l}),\eqn
$$
where the vectors {\bf l} range over a $d$-dimensional lattice
The reduction of multiple sums to products of single sums, when accomplished, provides an  elegant decrease in complexity. If this can be done we shall refer to the multiple sum as having been solved 'exactly'. There are many examples of this in the literature,where two and higher dimensional lattice sums may be solved exactly in terms of Dirichlet $L$-series,(Glasser and Zucker 1980;Zucker and Robertson 1984;Zucker 1990) but in this communication our concern is with two dimensional sums for which much more is known.  The classic example in this case, often ascribed to (Hardy 1920) but first given by (Lorenz 1871) is
$$
\sum_{m,n\neq 0,0}^{\infty}\f{1}{\left(m^2+n^2\right)^s}=
4\left(1+\f{1}{2^s}+\f{1}{3^s}+\f{1}{4^s}\ldots\right)\left(1-\f{1}{3^s}+\f{1}{5^s}-\f{1}{7^s}\ldots\right).\eqn
$$
The first series on the RHS of (1,2) represents the well-known Riemann zeta function
for $s>1$.  The second series is also well-known and often referred to as the Catalan beta function. These are both examples of series
introduced by Dirichlet in order to prove a famous theorem in number theory,
namely if $k$ and $l$ are relatively prime integers then the arithmetic
progression $kn+l$ contains an infinite number of primes.    Here we first survey the properties of $L$-series
which are relevant to lattice sums discussed below.

Elementary \L (modulo $k$) are given by
$$
L_k(s:x)=\s\f{\rchi_{k}(n)x^n}{n^s},\eqn
$$
%a1.1
$k$ will be referred to as the {\it period} or {\it order }of the
\L. Usually we are only concerned with $x=1$, and we write
$L_k(s:1):=L_k(s).$ In (1.3)  $\ch{k}$ is called a {\it character}.
It is a multiplicative function and periodic in $k$ and is defined
as follows. For $m,n$ integers
$$\eqalignno{
&\rchi_{k}(1)=1,\qquad \rchi_{k}(n+k)=\rchi_{k}(n),\qquad
\rchi_{k}(mn)=\rchi_{k}(m)\rchi_{k}(n),\cr &\rchi_{k}(n)=0\qquad
{\rm if}\ k\ {\rm and}\ n\ {\rm have\ a\ common\ factor} \neq
1.&\eq\cr}
$$
%1.4
It was shown by (Dickson 1939) that $\rchi_{k}$ can only assume
values which are the $\phi(k)$'th roots of unity, where $\phi(k)$ is the Euler
totient function which gives the number of positive integers less than $k$ which
are relatively prime to $k$.  Thus characters may be real or complex.  The properties of $L$-series with real characters are well-known and have been given elsewhere e.g. 
(Ayoub 1963; Zucker and Robinson 1976).  However the properties of $L$-series with complex characters seem less widely known and their properties are described here.
First, for completeness, we summarise the properties of real character $L$-series.
%Start anew section
\nexteq=1 \advance\sect by 1

\centerline {\bf 2.Properties of $L$-series with real characters.}

All $L$-series are divided into two types according to whether  $\rchi_{k}(k-1)= \pm 1$.  If  $\rchi_k(k-1)=+1$ the series will be said to have positive parity.  For such series the {\it signs} of the characters will be mirrored exactly in the midpoint of the series.  If  $\rchi_{k}(k-1)=-1$ the signs of the characters will be mirrored with opposite signs.
For real characters it may be shown that $\rchi_{k}=\pm 1$. Theorems
apposite to such $\rchi_{k}$ and given in (Ayoub 1963) show that the only possible characters are given by the
{\it Legendre-Jacobi-Kronecker} symbol, $\ch{k}(n) =(n\vert k)$.
The number of independent real character $L$-series is then found to
be as follows. Let $P=1$ or let $P$ be a product of all {\it
different} primes, i.e. $P$ is odd and square free, then
 \nl(a) if $k=P$  there is just one primitive \L.
 \nl(b) if $k=4P$ there is just one primitive \L.
 \nl(c) if $k=8P$ there are two primitive \L.
 \nl(d) if $k=2P$, $2^{\alpha}P$  where $\alpha>3$ , or $P$ is not square free there are \nl\phantom z\phantom z\phantom z\phantom z{\it no} primitive \L. \hfill\eq
 % 2.1
\nl The parity of a real $L$-series is determined as follows: \nl If
$k=P\equiv$1(mod4) the $L$-series has positive parity. \nl If
$k=P\equiv$3(mod4) the $L$-series has negative parity. \nl If $k=4P$
with $P\equiv$3(mod4) the $L$-series has positive parity. \nl If
$k=4P$ with $P\equiv$1(mod4) the $L$-series has negative parity. \nl
If $k=8P$ there is an $L$-series of each type. \nl The suffix $k$
will be {\it signed} according to whether $\rchi_{ k}(k-1)= \pm 1$.
If $s$ is a positive integer then explicit formulae for
$L_{-k}(2s-1)$ and $L_k(2s)$ may be established.  They are
$$
L_{-k}(2s-1)=\f{(-1)^{s-1}2^{2s-2}\pi^{2s-1}}{\sqrt{k}}
\sum_{n=1}^{k}\rchi_{-k}(n)\f{B_{2s-1}(1-n/k)}{(2s-1)!},
\eqn
$$
and
$$
L_{k}(2s)=\f{(-1)^{s-1}2^{2s-1}\pi^{2s}}{\sqrt{k}}
\sum_{n=1}^{k}\rchi_{k}(n)\f{B_{2s}(1-n/k)}{(2s)!},
\eqn
$$
where $B_s(x)$ are the Bernoulli polynomials. As both $n$ and $k$
are positive integers, and $B_s(1-n/k)$ are rational numbers, then
for $s$ a positive integer
$$\eqalignno{
L_{-k}(2s-1)=& R(k)\sqrt{k}\,\pi^{2s-1},&\eq\cr
L_{k}(2s)=& R'(k)\sqrt{k}\,\pi^{2s},&\eq\cr}
$$
%A1.14,1.15
where $R(k)$ and  $R'(k)$ are rational numbers.
It is also known that
$$
L_{k}(1)=\f{2h(k)}{\sqrt{k}}\log \epsilon_0,\eqn
$$
where $h(k)$ is the class number of the binary quadratic form of
discriminant $k$ and $\epsilon_0$ is the fundamental unit of the number
field $Q(\sqrt{k})$.  Some examples of real character $L$-series, which are relevant to this report are given below.  In order to do this concisely and further to emphasize the periodicity with which the series repeat themselves we introduce the notation
$$
(k,l:s):=(k,l)=\sum_{n=0}^{\infty}\f{1}{(kn+l)^s}.\eqn
$$
Thus the Riemann zeta function may be depicted as
$$
\zeta(s)=L_1(s)=\left(1+\f{1}{2^s}+\f{1}{3^s}+\f{1}{4^s}\ldots\right)=(1,1).
$$
The Catalan $\beta$ function is thus
$$
\beta(s)=L_{-4}(s)=\left(1-\f{1}{3^s}+\f{1}{5^s}-\f{1}{7^s}\ldots\right)=(4,1)-(4,3).
$$
Others which will be used here are
$$\displaylines{
L_{-3}(s)=(3,1)-(3,2),\cr
L_5(s)=(5,1)-(5,2)-(5,3)+(5,4),\cr
L_{-7}(s)=(7,1)+(7,2)-(7,3)+(7,4)-(7,5)-(7,6),\cr
L_{-8}(s)=(8,1)+(8,3)-(8,5)-(8.7),\cr
L_{8}(s)=(8,1)-(8,3)-(8,5)+(8.7),\cr
L_{12}(s)=(12,1)-(12,5)-(12,7)+(12,11),\cr
L_{-20}(s)=(20,1)+(20,3)+20,7)+(20,9)-(20,11)-(20,13)-(20,17)-(20,19),\cr
L_{28}(s)=(28,1)+(28,3)-(28,5)+(28,9)-(28,11)-(28,13)\cr
\hfill-(28,15)-(28,17)+(28,19)-(28,23)+(28,25)+(28,27).\hfill\eq\cr}
$$
%2.7
In addition to these there is always one further real \L of period $k$ formed by all the $(k,l)$ symbols being positive, but this can be shown to be equal to $
\Pi (1-p_n^{-s})L_1(s)$ where $p_n$ are all the different prime factors of $k$. This is illustrated  below for $k=5$ by making use of the expansion property of $(k,l)$ thus
$$
(k,l)=(2k,l)+(2k,k+l)=(3k,l)+(3k,k+l)+(3k,2k+l)\ldots.
$$
So
$$
(1,1)=(5,1)+(5,2)+(5,3)+(5,4)+(5,5)
$$
But $(5,5) = 5^{-s}(1,1)$ and
$$
(5,1)+(5,2)+(5,3)+(5,4)=(1,1)-(5,5)=(1-5^{-s})L_1
$$
Sums of the form
$$
Q(a,b,c;s):=Q(a,b,c)=\sum_{(m,n\neq 0,0)}^{\infty}(am^2+bmn+cn^2)^{-s}\eqn
$$
have been found exactly using real $L$-series.  Some examples are
$$\displaylines{
Q(1,1,2)=2L_1L_{-7},\quad Q(1,0,5)=L_1L_{-20}+L_{-4}L_5,\cr \hfil 2Q(5,2,10)=\left(1+7^{1-2s}\right)L_1L_{-4}-L_{-7}L_{28}.\hfil\eq\cr}
$$
In particular many $Q(1,0,\lambda)$ were solved by expressing them
as the Mellin transforms of the product of the Jacobian
$\theta_3(q)$ functions with different arguments: thus
$$\eqalignno{
Q(1,0,\lambda)=&\f{1}{\Gamma(s)}\if t^{s-1}\sum_{m,n\neq 0,0}^{\infty}\exp -\left[m^2 t+\lambda n^2 t\right]\d t\cr
=&\f{1}{\Gamma(s)}\if t^{s-1}\left[\theta_3(q)\theta_3(q^{\lambda})-1\right]\d t,\qquad {\rm where }\quad e^{-t}=q&\eq\cr
}
$$
%2.11
and
$$
\theta_3(q)=\sum_{n=-\infty}^{\infty}q^{n^2}=1+2q^2+2q^4+2q^9\ldots\eqn
$$
%2.12

By expressing $\theta_3(q)\theta_3(q^{\lambda})-1$ as Lambert series
where possible, the integral in (2.11) is easily evaluated in terms
of \L.  A large list of such results may be found in( Glasser and
Zucker 1980) supplemented by others in (Zucker and
Robertson 1984; Zucker 1990) In recent work - (McPhedran et.al. 2007) -
related to the Green's function connected with sums over the square
lattice, displaced lattice sums of the form
$$S(p,r,j;s)=S(p,r,j)=
{\sum_{m,n}} \left[\left(m+\f{p}{j}\right)^2
+\left(n+\f{r}{j}\right)^2\right]^{-s}\eqn
$$
were encountered. There were also associated phased lattice sums:
$$\sigma(p,r,j;s)=\sigma(p,r,j)=
{\sum_{(m,n\neq 0,0)}} \f{\exp[2\pi i(m p/j+n
r/j)]}{(m^2+n^2)^{s}},\eqn
$$
with the displaced and phased lattice sums connected by the Poisson
summation formula.

 It was found possible to solve $S(p,r,j)$ for all $j =2 - 4$ in
terms of real $L$-series, but for $j=5$ no such similar solutions,
apart from $S(1,2,5)$ and $S(1,3,10)$, were found. This prompted us to look for
solutions in terms of complex character $L$-series. Whereas the
properties of real \L are well known, complex \L  seem less well
documented, and in the next section some account of them is given.
%Start anew section
\nexteq=1 \advance\sect by 1

\centerline  {\bf 3.Properties of $L$-series with complex
characters.}

We first list all the possible \L with both real and complex characters for periods $k=1 - 10$ and for $k=16$.  For $k$ a prime these are found by taking each $\p(k)^{th}$ root of unity in turn  and using the rules for characters given in (1.4) to produce a given series always starting with +1 for the first term.  If $k$ has factors then these factors also have period $k$ and will reduce the number of independent \L of period $k$.  This will beome evident from the following.
$$k=1,\quad\phi(1)=1\qquad
(1,1)=L_1.
$$
$$k=2,\quad\phi(2)=1 \qquad  (2,1)=(1-2^{-s})L_1.$$

$k=3,\quad\phi(3)=2$.  1 and 2 being relatively prime to 3, there
are two possible order 3 L-series. \vskip -3em
$$
(3,1)+(3,2)=(1-3^{-s})L_1,\qquad (3,1)-(3,2)=L_{-3}.
$$
$k=4,\quad\phi(4)=2$ , 1 and 3 being relatively prime to 4,thus two
possible order 4 L-series exist. \vskip -3em
$$
(4,1)+(4,3)=(1-2^{-s})L_1,\qquad (4,1)-(4,3)=L_{-4}.
$$
$k=5,$ $\phi(5)=4$ and the four roots of unity are $\pm 1$ and $\pm
i$, giving
$$\eqalignno{
(5,1)+&(5,2)+(5,3)+(5,4)=(1-5^{-s})L_1.\cr
(5,1)-&(5,2)-(5,3)+(5,4)=L_5.\cr
(5,1)+&i(5,2)-i(5,3)-(5,4)=L_{-5}^{i}.\cr
(5,1)-&i(5,2)+i(5,3)-(5,4)=L_{-5}^{-i}.&\eq\cr\cr
}
$$
%3.1
 $k=6,$ $\phi(6)$=2, 1 and 5 are relatively prime to 6.  There are just two possible L-series both
 real:
$$\eqalignno{
&(6,1)+(6,5)=(1-2^{-s})(1-3^{-s})L_1.\cr &(6,1)-6,5)=(1+2^{-s})L_{-3}.&\eq\cr
}
$$
$k=7,$ $\phi(7)=6$.  1,2,3,4,5,6 being relatively prime to 7. The
sixth roots of unity are $1,-1,\omega,-\omega,\omega^2,-\omega^2$
where $\omega=\exp(i\pi/3)$.  The six possible L-series are
$$\eqalignno{
&(7,1)+(7,2)+(7,3)+(7,4)+(7,5)+(7,6)=(1-7^{-s})L_1,\cr
&(7,1)+(7,2)-(7,3)+(7,4)-(7,5)-(7,6)=L_{-7},\cr
&(7,1)+\ww(7,2)+\w(7,3)-\w(7,4)-\ww(7,5)-(7,6)=L_{-7}^{\ww},\cr
&(7,1)-\w(7,2)-\ww(7,3)+\ww(7,4)+\w(7,5)-(7,6)=L_{-7}^{-\w},\cr
&(7,1)+\ww(7,2)-\w(7,3)-\w(7,4)+\ww(7,5)+(7,6)=L_{7}^{\ww},\cr
&(7,1)-\w(7,2)+\ww(7,3)+\ww(7,4)-\w(7,5)+(7,6)=L_{7}^{-\w}.&\eq\cr }
$$

$\phi(8)=4$.  1,3,5,7 being relatively prime to 8, the four
functions are
$$\eqalignno{
&(8,1)+(8,3)+(8,5)+(8,7)=(1-2^{-s})L_1,\cr
&(8,1)-(8,3)+(8,5)-(8,7)=L_{-4},\cr
&(8,1)+(8,3)-(8,5)-(8,7)=L_{-8},\cr
&(8,1)-(8,3)-(8,5)+(8,7)=L_{8}.&\eq\cr }
$$
$\phi(9) =6$.  1,2,4,5,7,8 being relatively prime to 9, there are
six functions:
$$\eqalignno{
&(9,1)+(9,2)+(9,4)+(9,5)+(9,7)+(9,8)=(1-3^{-s})L_{1},\cr
&(9,1)-(9,2)+(9,4)-(9,5)+(9,7)-(9,8)=L_{-3},\cr
&(9,1)-w^2(9,2)-w(9,4)+w(9,5)]+w^2(9,7)-(9,8)=L_{-9}^{-\ww},\cr
&(9,1)+w(9,2)+w^2(9,4)-w^2(9,5)-w(9,7)-(9,8)=L_{-9}^{\w},\cr
&(9,1)+w^2(9,2)-w(9,4)-w(9,5)+w^2(9,7)+(9,8)=L_{9}^{\ww},\cr
&(9,1)-w(9,2)+w^2(9,4)+w^2(9,5)-w(9,7)+(9,8)=L_{9}^{-\w}.&\eq\cr}
$$
$\phi(10)=4$.  1,3,7,9 being relatively prime to 10, the four
functions are
$$\eqalignno{
(10,1)+&(10,3)+(10,7)+(10,9)=(1-2^{-s})(1-5^{-s})L_1,\cr
(10,1)-&(10,3)-(10,7)+(10,9)=(1+2^{-s})L_{5},\cr
(10,1)-&i(10,3)+i(10,7)-(10,9)=(1-i2^{-s})L_{-5}^{i},\cr
(10,1)+&i(10,3)-i(10,7)-(10,9)=(1+i2^{-s})L_{-5}^{-i}&.\eq \cr}
$$
$\phi(16)=8$. 1,3,5,7,9,11,13,15 being relatively prime to 16, there
are eight functions.  As $\phi(k)$ gets larger the depiction of \L in  $(k,l)$ symbols can become unwieldy. It is thus convenient here to introduce a slightly modified notation namely a {\it signed} $(k,l)$ symbol defined by.
$$(k,l)_+:=(k,l)+(k,k-l),~~(k,l)_-:=(k,l)-(k,k-l).$$
This enables us to halve the number of symbols required to show an $L$-series.  It will be seen that positive parity series are described entirely by $(k,l)_+$ symbols and negative parity series by $(k,l)_-$ terms.  Thus for $k=16$ we have
$$\displaylines{
(16,1)_++(16,3)_++(16,5)_++(16,7)_+=(1-2^{-s})L_1,\cr
(16,1)_--(16,3)_-+(16,5)_--(16,7)_-=L_{-4},\cr
(16,1)_-+(16,3)_--(16,5)_--(16,7)_-=L_{-8},\cr
(16,1)_+-(16,3)_+-(16,5)_++(16,7)_+=L_8,\cr
(16,1)_-+i(16,3)_-+i(16,5)_-+(16,7)_-=L_{-16}^{i},\cr
(16,1)_--i(16,3)_--i(16,5)_-+(16,7)_-=L_{-16}^{-i},\cr
(16,1)_++i(16,3)_+-i(16,5)_+-(16,7)_+=L_{16}^{i},\cr
\hfil (16,1)_+-i(16,3)_++i(16,5)_+-(16,7)_+=L_{16}^{-i}.\hfil\eq}
$$
These results illustrates most of the properties that have been
observed  with complex \L.  Since for $k>2,\,\,\phi(k)$ is always even \nl (a) The number of positive parity
series always  equals the number of negative parity ones. \nl (b)The number of complex character positive parity \L is even and they divide into pairs of complex conjugates.  The same is true for complex character \L with negative parity.\nl(c)
The first and the last terms of all \L are always real. \nl (d) Knowing the
parity of the series and the first non-real term then using the
properties of characters given in (1.4) the coefficients of all the
other terms may be established.  It allows us to establish a concise
notation for complex \L .  The subscript gives the parity and
period, whilst the superscript gives the coefficient of the first
non-real term, and this specifies the given series completely.  \nl
(e) The number of $(k,l)$ symbols needed to specify a given \L
equals the number of \L for a given $k$.  Thus every $(k,l)$ is
expressible as a linear combination of \L of period $k$.

We note from the above that for $k=6, 10$ the inclusion of complex \L do not yield any \L with these periods, and this has been found for $k=14$ and 18.  So part of the statement (2.1) (d) namely if $k$ is equal to twice an odd number no \L of such a period exists. It would appear that this is the result of the fact that $\phi(2n)=\phi(n)$ if $n$ is an odd number.   However, whereas for real \L if $k$ is a perfect square $>4$ no such period \L are found, it is seen here that complex \L for such $k$ do exist.

It is necessary to point out here that all the statements made about real \L made in the previous section have been proved, whereas the assertions made in this section about complex \L, though we believe them to be entirely true, have not been proved.
%Start anew section
\nexteq=1 \advance\sect by 1
\centerline{\bf 4.Expressions for  $S(p,r,j)$ in closed form for $j=2-10$.}

It is easily seen that $S(p,r,j)=S(j-p,j-r,j)$, so all the independent $S$ for a given $j$ will be found by allowing both $p$ and $r$ to take on all values up to and including $j/2$.  Then the displaced lattice sums may be expressed as the following Mellin transform.
$$\displaylines{
S(p,r,j)= {\sum_{m,n}} \f{1}{\left[\left(m+\f{p}{j}\right)^2
+\left(n+\f{r}{j}\right)^2\right]^s}={\sum_{m,n}}
\f{j^{2s}}{\left[\left(jm+p\right)^2
+\left(jn+r\right)^2\right]^s}\cr \hfill=\f{j^{2s}}{\Gamma(s)}\if
t^{s-1}\sum_{m,n-\infty}^{\infty}\exp -\left[(jm+p)^2 t+(jn+r)^2
t\right] \d t.\hfill\eq\cr}
$$
The exponential sums are disjoint and each can be evaluated
separately. Then letting $e^{-t}=q$ and writing
$$
\sum_{m=-\infty}^{\infty}q^{(jm+p)^2}=\theta(j,p)\eqn
$$
we have
$$
S(p,r,j)=\f{j^{2s}}{\Gamma(s)}\if t^{s-1}\theta(j,p)\theta(j,r)\d t.\eqn
$$
For small $j$,  $\theta(j,p)$ may be expressed in terms of  $\theta_3 $-functions - [{\bf 9}], and we list the results for $j=2,3,4,6$.
$$\displaylines{
\theta(j,0)=\theta_3\left(q^{j^2}\right),\quad \theta(2,1)=\theta_3(q)-\theta_3\left(q^4\right),
\quad \theta(3,1)=\f{1}{2}\left[\theta_3(q)-\theta_3\left(q^9\right)\right]\cr\hfill
\theta(4,1)=\f{1}{2}\left[\theta_3(q)-\theta_3(q^4)\right],\quad\theta(4,2)=\theta_3(q^4)-\theta_3\left(q^{16}\right).\hfill\eq\cr
}
$$
$$\eqalignno{
\theta(6,0)=&\theta_3(q^{36}),\qquad
\theta(6,1)=\f{1}{2}\left[\theta_3(q)-\theta_3(q^4)-\theta_3(q^9)+\theta_3(q^{36})\right],\cr
\theta(6,2)=&\f{1}{2}\left[\theta_3(q^4)-\theta_3(q^{36})\right],\qquad
\theta(6,3)=\theta_3(q^9)-\theta_3(q^{36}),&\eq\cr}
$$
%
%1.5
For these $j$ values all the independent $S(p,r,j)$ may be expressed in terms of $Q(1,0,\lambda)$ using (2.11) thus
$$\displaylines{
S(0,1,2)=2^{2s}\left[Q(1,0,4)-4^{-s}Q(1,0,1)\right]\cr
S(1,1,2)=2^{2s}\left[Q(1,0,1)-2Q(1,0,4)+4^{-s}Q(1,0,1)\right]\cr
S(0,1,3)=\f{3^{2s}}{2}\left[Q(1,0,9)-9^{-s}Q(1,0,1)\right]\cr
S(1,1,3)=\f{3^{2s}}{2}\left[Q(1,0,1)-2Q(1,0,9)+9^{-s}Q(1,0,1)\right]\cr
S(0,1,4)=\f{4^{2s}}{2}\left[Q(1,0,16)-4^{-s}Q(1,0,4)\right]\cr
S(0,2,4)=\f{4^{2s}}{2}\left[2^{-2s}Q(1,0,4)-2^{-4s}Q(1,0,1)\right]\cr
S(1,1,4)=\f{4^{2s}}{2}\left[Q(1,0,1)-2Q(1,0,4)+4^{-s}Q(1,0,4)\right]\cr
S(1,2,4)=\f{4^{2s}}{2}\left[Q(1,0,4)-4^{-s}Q(1,0,1)-Q(1,0,16)-4^{-s}Q(1,0,4)\right]\cr
S(0,0,6)=Q(1,0,1).\cr
S(0,1,6)=\f{1}{2}
\left[6^{-2s}Q(1,0,1)-3^{-2s}Q(1,0,4)-2^{-2s}Q(1,0,9)+Q(1,0,36)\right].\cr
S(1,1,6)=\f{6^{2s}}{4}
\bigl[(1+2^{-2s}+3^{-2s}+6^{-2s})Q(1,0,1)\cr
-2(1+3^{-2s})Q(1,0,4)-2(1+2^{-2s})Q(1,0,9)
+2Q(4,0,9)+2Q(1,0,36)\bigl].\cr
S(0,2,6)=\f{3^{2s}}{2}\left[-3^{-2s}Q(1,0,1)+Q(1,0,9)\right].\cr
S(1,2,6)=\f{6^{2s}}{4}
\bigl[-2^{-2s}(1+3^{-2s})Q(1,0,1)+(1+3^{-2s})Q(1,0,4)\cr
+2^{1-2s}Q(1,0,9)-Q(4,0,9)-Q(1,0,36)\bigl].\cr
S(2,2,6)=\f{3^{2s}}{2}\left[(1+3^{-2s})Q(1,0,1)-2Q(1,0,9)\right].\cr
S(0,3,6)=-Q(1,0,1)+2^{-2s}Q(1,0,4).\cr
S(1,3,6)=\f{6^{2s}}{2}
\bigl[(-3^{-2s}-6^{-2s})Q(1,0,1)+2.3^{-2s}Q(1,0,4)\cr
+(1+2^{-2s})Q(1,0,9)
-Q(4,0,9)-Q(1,0,36)\bigl].\cr
S2,3,6)=\f{6^{2s}}{2}
\left[6^{-2s}Q(1,0,1)-3^{-2s}Q(1,0,4)-2^{-2s}Q(1,0,9)+Q(4,0,9)\right].\cr
\hfill S(3,3,6)=2^{2s}\bigl[(1+2^{-2s})Q(1,0,1)-2Q(1,0,4)\bigl]\hfill\eq
\cr}
$$
Of the six different $Q(a,b,c)$ which appear in the preceding displaced sums, the values of four have been found in terms of Dirichlet series with real characters. They are
$$\displaylines{
 Q(1,0,1)=4L_1(s)L_{-4}(s),\qquad Q(1,0,4)=2(1-2^{-s}+2^{1-2s})L_1(s)L_{-4}(s),\cr Q(1,0,9)=(1+3^{1-2s})L_1(s)L_{-4}(s)+L_{-3}(s)L_{12}(s).\cr
 \hfill Q(1,0,16)=(1-2^{-s}+2^{1-2s}-2^{1-3s}+2^{2-4s})L_1(s)L_{-4}(s)+L_{-8}(s)L_8(s).\hfill\eq\cr}
$$
Q(1,0,36) and Q(4,0,9) cannot be individually found.  However, it will be seen that
$\pm\left[Q(1,0,36)+Q(4,0,9)\right]$ appear together in three of the cases considered,and it may be shown via the theory of which numbers the binary quadratic forms $(m^2+36n^2)$ + $(4m^2+9n^2)$ represent that
$$\displaylines{ Q(1,0,36)+Q(4,0,9)=(1-2^{-s}+2^{1-2s})(1+3^{1-2s})L_1(s)L_{-4}(s)\cr\hfill +(1+2^{-s}+2^{1-2s})L{_3}(s)L_{12}(s)\dagger.\hfill\eq\cr}
$$
$\dagger$(We are grateful to Mark Watkins,University of Bristol, for obtaining this result for us.)

So we are able to express eight of the ten possible $S(p,r,6)$ in terms of known Dirichlet series.  All the results for $j=2-4$ and $j=6$ are listed below
$$\displaylines{
j=2,\quad S(0,1,2)=2^{2s+1}\ll.\quad S(1,1,2)=2^{s+2}\ll.\cr
j=3,\hfill S(0,1,3)=\f{3^{2s}}{2}\left[(1-3^{-2s})L_1(s)L_{-4}(s)+L_{-3}(s)L_{12}(s)\right].\hfill\cr
S(1,1,3)=\f{3^{2s}}{2}\left[(1-3^{-2s})L_1(s)L_{-4}(s)-L_{-3}(s)L_{12}(s)\right].\cr
j=4,\hfill S(0,1,4)=\f{4^{2s}}{2}\left[\ll+L_{-8}(s)L_8(s)\right].\hfill\cr
S(0,2,4)=2^{2s+1}\ll.\cr S(1,1,4)=2^{3s}\ll.\cr
S(1,2,4)=\f{4^{2s}}{2}\left[\ll-L_{-8}(s)L_8(s)\right],\cr
j=6,\hfill
S(0,1,6)=\f{1}{2}\bigl[(-3+2^{1+s}-2^{1+2s}-3^{2s})L_1(s)L_{-4}(s)\hfill\cr
-3^{2s}L_{-3}(s)L_{12}(s)+6^{2s}Q(1,0,36)\bigr].\cr 
S(1,1,6)=\f{1}{2}
\left[(2^{s}-1)(3^{2s}-1)L_1(s)L_{-4}(s)+3^{2s}(2^{s}+1)L_{-3}(s)L_{12}(s)\right]\cr
S(0,2,6)=\f{1}{2}\left[(3^{2s}-1)L_1(s)L_{-4}(s)+3^{2s}L_{-3}(s)L_{12}(s)\right].\cr
S(1,2,6)=\f{1}{4}
\left[2^{s}(2^{s}+1)(3^{2s}-1)L_1(s)L_{-4}(s)-18^{s}(2^s+1)L_{-3}(s)L_{12}(s)\right].\cr
S(2,2,6)=\f{1}{2}\left[(3^{2s}-1)L_1(s)L_{-4}(s)-3^{2s}L_{-3}(s)L_{12}(s)\right].\cr
S(0,3,6)=2^{1+s}(2^{s}-1)L_1(s)L_{-4}(s).\cr S(1,3,6)=\f{1}{2}
\left[(2^{s}-1)(3^{2s}-1)L_1(s)L_{-4}(s)-3^{2s}(2^{s}+1)L_{-3}(s)L_{12}(s)\right]\cr
S(2,3,6)=
\f{1}{2}\bigl[(-3+2^{1+s}-2^{1+2s}-3^{2s})L_1(s)L_{-4}(s)\cr
-3^{2s}L_{-3}(s)L_{12}(s)+6^{2s}Q(4,0,9)\bigr].\cr 
 \hfill S(3,3,6)=4(2^{s}-1)L_1(s)L_{-4}(s).\hfill\eq\cr }
$$
It may also be seen that the sum of the two unknown members may also be expressed
in terms of Dirichlet series, thus
$$
S(0,1,6)+S(2,3,6)=
2^{s-1}(2^s-1)(3^{2s}-1)L_1(s)L_{-4}(s)+2^{s-1}3^{2s}(2^s+1)L_{-3}(s)L_{12}(s)
$$
Whether or not $Q(1,0,36)$ or $Q(4,0,9)$ may be individually
expressed in terms of Dirichlet series with complex characters is
still an open question.

For $j=5$, the previous method can be used to establish one
factorization,(Zucker 1990) :
$$
S(1,2,5)= (5^{s}-1)L_1(s)L_{-4}(s).\eqn
$$
The other results for this case all involve Dirichlet series with
complex characters, and have been obtained (McPhedran et al.2007))
by evaluating  coefficients $c_n$ in the expansion
$S(k,l,5)=\sum_{n=1}^\infty \f{c_n}{n^s}$ for a sufficient set of
$c_n$. These may be compared with the corresponding expansions
generated from appropriate combinations of Dirichlet functions,
where if a product with a pair of complex characters occurs, this
must be accompanied by the product with complex conjugated
characters, to ensure a real result for $\Im (s)=0$.
We need the
following two complex $L$ functions of order 20:
$$\displaylines{
L_{20}^i(s)=(20,1)_++i(20,3)_+-i(20,7)_+-(20,9)_+,\cr
\hfill\qquad\quad L_{20}^{-i}(s)=(20,1)_+-i(20,3)_++i(20,7)_+-(20,9)_+.\hfill\eq\cr}$$
and writing
$$\displaylines{
a5(s)=(1-5^{-s})^2L_1(s)L_{-4}(s)-L_5(s)L_{-20}(s)\cr
b5(s)=(1-5^{-s})^2L_1(s)L_{-4}(s)+L_5(s)L_{-20}(s)\cr
c51(s)=L_{-5}^i(s)L_{20}^i(s)+L_{-5}^{-i}(s)L_{20}^{-i}(s)\cr
c52(s)=i\left[L_{-5}^i(s)L_{20}^i(s)-L_{-5}^{-i}(s)L_{20}^{-i}(s)\right]
}
$$
then we have
$$\displaylines{
S(0,1,5)=\f{5^{2 s}}{4}[b5(s)+c51(s)],\cr
S(1,1,5)=\f{5^{2 s}}{4}[a5(s)-c52(s)],\cr
S(0,2,5)=\f{5^{2s}}{4}[b5(s)-c51(s)],\cr
\hfill \qquad\quad S(2,2,5)=\f{5^{2s}}{4}[a5(s)+c52(s)],\hfill\eq
\cr }
$$

Note that in  the expansion of $S(p,r,j)$ as a sum over factors $1/n^s$, all terms with non-zero coefficients
have $n=(p^2+r^2) mod (j)$. For $j=5$, each modulus value 1,2,3,4,5 occurs only once.

We are now able to append the following results. For $j=7$, there are nine lattice sums to be evaluated, of which we have found expressions for three, and pair relations connecting the other six.
 The new complex $L$ functions we will need are of order 28 and are:
$$\displaylines{
(28,1)_-+\w(28,3)_-+\ww(28,5)_-+\ww(28,9)_-+\w(28,11)_-+(28,13)_-=L_{-28}^{\w}\cr
(28,1)_--\ww(28,3)_--\w(28,5)_--\w(28,9)_--\ww(28,11)_-+(28,13)_-=L_{-28}^{-\ww}\cr
(28,1)_+-\w(28,3)_+-\ww(28,5)_++\ww(28,9)_++\w(28,11)_+-(28,13)_+=L_{28}^{-\w}\cr
(28,1)_++\ww(28,3)_++\w(28,5)_+-\w(28,9)_+-\ww(28,11)_+-(28,13)_+=L_{28}^{\ww},
\cr}
$$
Then define
$$\displaylines{
a7(s)=(1-7^{-2s})L_1(s)L_{-4}(s)-L_{-7}(s)L_{28}(s)\cr
b7(s)=(1-7^{-2s})L_1(s)L_{-4}(s)+L_{-7}(s)L_{28}(s)\cr
c71(s)=L_7^{\w^2}(s) L_{-28}^{\w}(s)+L_7^{-\w}(s)L_{-28}^{-\w^2}(s)\cr
c72(s)=L_{-7}^{\w^2}(s)L_{28}^{-\w}(s)+L_{-7}^{-\w}(s)L_{-28}^{\w^2}(s)\cr
c73(s)=\w L_7^{\w^2}(s)L_{-28}^{\w}(s)-\w^2L_7^{-\w}(s)L_{-28}^{-\w^2}(s)\cr
c74(s)=\w L_{-7}^{\w^2}(s)L_{28}^{-\w}(s)-\w^2L_{-7}^{-\w}(s)L_{-28}^{\w^2}(s)\cr
c75(s)=\w^2 L_7^{\w^2(s)}L_{-28}^{\w}(s)-\w L_7^{-\w}(s)L_{-28}^{-\w^2}(s)\cr
\hfill c76(s)=\w^2 L_{-7}^{\w^2}(s)L_{28}^{-\w}(s)-\w L_{-7}^{-\w}(s)L_{-28}^{\w^2}(s)\hfill\eq\cr
}
$$
The three sums which have been solved have expansions with terms with values 6,5 and 3(modulus 7) respectively , are
$$\displaylines{
S(2,3,7)=\f{7^{2s}}{12}\left[a7(s)+c71(s)-c72(s)\right]\cr
S(1,2,7)=\f{7^{2s}}{12}\left[a7(s)-c73(s)+c74(s)\right]\cr
S(1,3,7)=\f{7^{2s}}{12}\left[a7(s)+c75(s)-c76(s)\right]\cr
}
$$
The three sets of pair relations contain terms with values of 1, 2 and 4 (mod 7)respectively,  and are
$$\displaylines{
S(0,1,7)+S(2,2,7)=\f{7^{2s}}{6}\left[b7(s)+c71(s)+c72(s)\right]\cr
S(0,3,7)+S(1,1,7)=\f{7^{2s}}{6}\left[b7(s)-c73(s)-c74(s)\right]\cr
S(0,2,7)+S(3,3,7)=\f{7^{2s}}{6}\left[b7(s)+c75(s)+c76(s)\right]\cr}
$$
For $j=8$,
let
$$\displaylines{
a8(s)=(1-2^{-s})L_1(s)L_{-4}(s)-L_{-8}(s)L_{8}(s)\cr
b8(s)=(1-2^{-s})L_1(s)L_{-4}(s)+L_{-8}(s)L_{8}(s)\cr
c81(s)=L_{-16}^{-i}(s)L_{16}^i(s)+L_{-16}^i(s)L_{16}^{-i}(s)\cr
c82(s)=i\left[L_{-16}^{-i}(s)L_{16}^i(s)-L_{-16}^i(s)L_{16}^{-i}(s)\right]\cr
}
$$
The following three individual sums have been found
$$\displaylines{
S(1,2,8)=2^{6s-3}\left[a8(s)+c82(s)\right]\cr
S(1,3,8)=2^{5s-2}a(s)\cr
S(2,3,8)=2^{6s-3}\left[a8(s)-c82(s)\right]\cr
}
$$
The six remaining independent sums occur as three pairs:
$$\displaylines{
S(0,1,8)+S(1,4,8)=2^{6s-2}\left[b8(s)+c81(s)\right]\cr
S(0,3,8)+S(3,4,8)=2^{6s-2}\left[b8(s)-c81(s)\right]\cr
S(1,1,8)+S(3,3,8)=2^{5s-1}b8(s)\cr
}
$$

For $j=9$ there are twelve independent terms. We require the following $L$ functions of order 36:
$$\displaylines{
(36,1)_--\w(36,5)_--\ww(36,7)_--\ww(36,11)_--\w(36,13)_-+(36,17)_-=L_{-36}^{-\w}\cr
(36,1)_-+\ww(36,5)_-+\w(36,7)_-+\w(36,11)_-+\ww(36,13)_-+(36,17)_-=L_{-36}^{\ww}\cr
(36,1)_++\w(36,5)_--\ww(36,7)_++\ww(36,11)_+-\w(36,13)_+-(36,17)_+=L_{36}^{\w}\cr
(36,1)_+-\ww(36,5)_-+\w(36,7)_+-\w(36,11)_++\ww(36,13)_+-(36,17)_+=L_{36}^{-\ww}\cr
}
$$
Then define
$$\displaylines{
a9(s)=(1-3^{-2s})L_1(s)L_{-4}(s)-L_{-3}(s)L_{12}(s)\cr
b9(s)=(1-3^{-2s})L_1(s)L_{-4}(s)+L_{-3}(s)L_{12}(s)\cr
c91(s)=L_9^{\w^2}(s)L_{-36}^{-\w}(s)+L_9^{-\w}(s)L_{-36}^{\w^2}(s)\cr
c92(s)=L_{-9}^{-\w^2}(s)L_{36}^{\w}(s)+L_{-9}^{\w}(s)L_{36}^{-\w^2}(s)\cr
c93(s)=\w L_9^{\w^2}(s)L_{-36}^{-\w}(s)-\w^2L_9^{-\w}(s)L_{-36}^{\w^2}(s)\cr
c94(s)=\w L_{-9}^{-\w^2}(s)L_{36}^{\w}(s)-\w^2L_{-9}^{\w}(s)L_{36}^{-\w^2}(s)\cr
c95(s)=\w^2 L_9^{\w^2}(s)L_{-36}^{-\w}(s)-\w L_9^{-\w}(s)L_{-36}^{\w^2}(s)\cr
c96(s)=\w^2L_{-9}^{-\w^2}(s)L_{36}^{\w}(s)-\w L_{-9}^{\w}(s)L_{36}^{-\w^2}(s)\cr
}
$$
 and we find the twelve sums divide into six pairs thus
$$\displaylines{
S(2,2,9)+2S(1,4,9)=\f{9^{2s}}{6}\left[(a9(s)+c91(s)-c92(s))\right]\cr
S(1,1,9)+2S(2,4,9)=\f{9^{2s}}{6}\left[(a9(s)-c93(s)+c94(s))\right]\cr
S(4,4,9)+2S(1,2,9)=\f{9^{2s}}{6}\left[(a9(s)+c95(s)-c96(s))\right]\cr
S(0,1,9)+2S(1,3,9)=\f{9^{2s}}{6}\left[(b9(s)+c91(s)+c92(s))\right]\cr
S(0,4,9)+2S(3,4,9)=\f{9^{2s}}{6}\left[(b9(s)-c93(s)-c94(s))\right]\cr
S(0,2,9)+2S(2,3,9)=\f{9^{2s}}{6}\left[(b9(s)+c95(s)+c96(s))\right]\cr
}
$$
These results as displayed are associated with terms having values 8,2,5,1,7,4 (mod)9 respectively.  The similarity between these results and those of $j=7$ is noteworthy.

For $j=10$, we need the complex $L$ functions of order 20, given for $j=5$.
We have the following definitions and expressions for sums:
$$\displaylines{
a(s)=(1-2^{-s})(1-5^{-s})^2L_1(s)L_{-4}(s)-(1+2^{-s})L_{5}(s)L_{-20}(s)\cr
b(s)=(1-2^{-s})(1-5^{-s})^2L_1(s)L_{-4}(s)+(1+2^{-s})L_{5}(s)L_{-20}(s)\cr
c(s)=L_{-5}^{i}(s)L_{20}^i(s)+L_{-5}^{-i}(s)L_{20}^{-i}(s)\cr
d(s)=i\left[L_{-5}^{i}(s)L_{20}^i(s)-L_{-5}^{-i}(s)L_{20}^{-i}(s)\right]\cr
}
$$
$$\displaylines{
S(1,1,10)=\f{10^{2s}}{2^{s+2}}\left[b(s)+c(s)-2^{-s}d(s)\right]\cr
S(1,3,10)=10^{s}(1-2^{-s})(1-5^{-s})L_1(s)L_{-4}(s)\cr
S(1,4,10)=\f{10^{2s}}{8}\left[a(s)-2^{-s}c(s)-d(s)\right]\cr
S(1,5,10)=\f{10^{2s}}{2^{s+2}}\left[a(s)+2^{-s}c(s)+d(s)\right]\cr
S(2,3,10)=\f{10^{2s}}{8}
\left[a(s)+2^{-s}c(s)+d(s)\right]\cr
S(3,3,10)=\f{10^{2s}}{2^{s+2}}\left[b(s)-c(s)+2^{-s}d(s)\right]\cr
S(3,5,10)=\f{10^{2s}}{2^{s+2}}\left[a(s)-2^{-s}c(s)-d(s)\right]\cr
S(0,1,10)+S(4,5,10)=\f{10^{2s}}{4}\left[b(s)+c(s)-2^{-s}d(s)\right]\cr
S(0,3,10)+S(2,5,10)=\f{10^{2s}}{4}\left[b(s)-c(s)+2^{-s}d(s)\right]\cr
S(1,2,10)+S(3,4,10)=\f{10^{2s}}{5^s}(1-2^{-s})(1-5^{-s})L_1(s)L_{-4}(s)\cr}
$$
It appears to us that there is no apparent reason why similar results cannot be found for larger values of $j$.  However,at this stage we have no rules for determining which particular $S(p,r,j)$ or combination of such terms can be put into closed form and some criteria are desirable in order to go further.

%Start anew section

\nexteq=1 \advance\sect by 1
\centerline {\bf 5. Exact solutions of a lattice sum involving an indefinite quadratic form.}

Efforts to solve $Q(a,b,c)$ in terms of \L have concentrated on the case when the binary quadratic form $am^2+bmn+cn^2$ is positive definite, i.e. $a>0$ and the discriminant $b^2-4ac<0$. Following Lorenz[{\bf 5}] who found an exact form for T(1,0 -1), Zucker and Robertson([{\bf 8}] attempted to solve some lattice sums involving indefinite quadratic forms.  They investigated
$$
\sum_{p^2m^2\neq r^2n^2}\vert p^2m^2-r^2n^2\vert ^{-s}=T(p^2,0,-r^2;s):=T(p^2,0,-r^2),\eqn
$$
%5.1
and in particular found an expression for $T(1,0,-r^2)$
in terms of $(k,l)$ symbols defined in (2.7).  This was
$$
T(1,0,-r^2)=4r^{-2s}\left(1-2^{1-s}+2^{1-2s}\right)L_1^2+2\sum_{t=1}^{r-1}\left[(2r,t)+(2r,2r-t)\right]^2.\eqn
$$
%5.2
For $r=1 - 6$ it was found possible to find solutions in terms of {\it squares} of positive parity real \L, but for larger values of $r$ no such solutions could be found.  Now it was noted in section 3 that  every $(k,l)$ is expressible as a linear combination of \L of period $k$ if the complex \L are included.  Thus in principal (5.2) may be written in \L for {\it every} $r$.  To illustrate this, closed forms for $r=1 - 13$ have been evaluated and the outcome displayed in Table 1.
In the table we have
$$
T(1,0,-r^2)=\sum_{m^2\neq r^2n^2}\f{1}{\vert m^2-r^2n^2\vert^s}\eqn
$$
and
$$
\omega=\exp\left(\f{i\pi}{3}\right),\qquad \tau=\exp\left(\f{i\pi}{5}\right),\qquad\p=\exp\left(\f{i\pi}{6}\right).\eqn
$$
These results show that in addition to squares of positive parity \L with real characters,products of pairs of positive parity
complex \L will in general be required to solve $T(1,0,-r^2)$. It is apparent that with sufficient labour there is no limit to which one may go, but until now no clear pattern has yet emerged.

 We have established numerically, but not analytically, the following functional equation:
 $$
 T(1,0,-r^2;s)=T(1,0,-r^2;1-s)\left(\f{\pi}{r}\right)^{2 s-1}\f{\Gamma(1-s)}{\Gamma(s)}\tan \left(\f{s\pi}{2}\right).\eqn
 $$
It may also be shown that $T$ has the following expansion near its second-order pole at $s=1$:
 $$
T(1,0,-r^2;1+s)=\f{2}{rs^2}+\f{4\gamma}{rs}+\f{2}{r}\left(\gamma^2+2\gamma_1\right)+C(r)+O[s].\eqn
$$
$\gamma_1$ is the first Stieltjes constant and $C(r)$ is a constant depending on $r$.  The first three are
$$
C(1)=2\log^2 2,\qquad C(2)=\f{3}{2}\log^2 2,\qquad C(3)=\f{2}{3}\log^2 2+\f{1}{3}\log^2 3.
$$
The expansion near $s=0$ is then
$$
T(1,0,-r^2;s)=1+2\log \left(\f{\pi}{r}\right) s+O(s^2).\eqn
$$
{\centerline {\bf Conclusion.}}
\nl It has been found that certain two dimensional lattice sums which hitherto were not expressible in closed form, may now be written exactly if \L with complex characters  are employed.  This increases considerably the number of two dimensional lattice sums which can be expressed in closed form.  An obvious question is can  complex character \L play a role in higher dimensional sums, and this may be an interesting path to pursue.
\nl{\bf Acknowledgment}
RCM acknowledges the support of the Australian Research Council, and valuable discussions with Prof. Lindsay Botten and Dr. Nicolae Nicorovici.

\vfill\break
{\bf References}
\nl Ayoub,R. 1963 {\it An Introduction to the Analytic Theory of Numbers.}  Providence:
American Mathematical Society .
\nl Dickson,L.E. 1939 {\it An Introduction to the Analytic Theory of
Numbers.} Chicago: University of Chicago Press.
\nl Glasser,M.L. and  Zucker,I.J. 1980 Lattice Sums. {\it
Theoretical Chemistry.Advances and Perspectives}. {\bf 5} , 67-139. New York,NY:Academic Press 
\nl Hardy,G.H. 1920 On some definite integral considered by Mellin.
{\it Mess.Math.}{\bf 49}, 86-91
\nl Lorenz,L. 1871  Bidrag tiltalienes theori, {\it Tidsskrift for Math.} {\bf 1},  97-114.
\nl McPhedran,R.C.,Botten,L.C.,Nicorovici,N.P. and Zucker,I.J. 2007
Systematic investigation of two-dimensional static array sums. {\it J. Math. Phys.} {\bf 48}  033501.
\nl Zucker,I.J. and Robertson,M.M. 1976 Some Properties of Dirichlet L- Series, {\it J. Phys. A.} {\bf 9} ,
1207-1214.
\nl Zucker,I.J. and Robertson,M.M. 1984 Further Aspects of the Evaluation of\nl 
$\sum_{(m,n,\neq 0,0)}(am^2+bmn+cn^2)^{-s}$.
{\it Math. Proc. Camb. Phil. Soc.} {\bf 95},  5-13.
\nl Zucker,I.J. 1990 Further Relations Amongst Infinite Series and Products. II. The Evaluation of 3-Dimensional Lattice Sums. {\it J. Phys. A.} {\bf 23}, 117-132.
\vfil\break
\centerline{ Table 1}
\phantom z
\centerline{
\vbox{\offinterlineskip  %Creates a vbox andstops the normal lineskip
\halign{\strut#&\vrule#&  %Replaces lineskip forminginvisible 1st column.
                         %\vrule# puts in the leftmost vertical rule.  %1st column followed by a vertical rule
\quad\hfil#\hfil&\quad\vrule#&  %2nd column followed by a vertical rule
\quad\hfil#\hfil&\quad\vrule#\cr  %3rd column followed by a vertical rule
                           %End of tenplate
\noalign{\hrule}           %Top horizontal line
\omit&height6pt&\omit&&\omit&\cr %Extra space
&& $r$
&&$T(1,0,-r^2)$ &\cr
\omit&height6pt&\omit&&\omit&\cr %Extra space
\noalign{\hrule}
\omit&height6pt&\omit&&\omit&\cr %Extra space
&& $1$
&& $4\left(1-2^{1-s}+2^{1-2s}\right)L_1^2(s) $ &\cr
\omit&height6pt&\omit&&\omit&\cr %Extra space
\noalign{\hrule}
\omit&height6pt&\omit&&\omit&\cr %Extra space
&& $2$
&& $2\left[1+\left(\sqrt{2}-1\right)2^{-s}+2^{1-2s}\right]\left[1-\left(\sqrt{2}+1\right)2^{-s}+2^{1-2s}\right]L_1^2(s)$ &\cr
\omit&height6pt&\omit&&\omit&\cr %Extra space
\noalign{\hrule}
\omit&height6pt&\omit&&\omit&\cr %Extra space
&& $3$
&& $ 2\left(1-2^{1-s}+2^{1-2s}\right)\left(1-2.3^{-s}+3^{1-2s}\right)L_1^2(s)$ &\cr
\omit&height6pt&\omit&&\omit&\cr %Extra space
\noalign{\hrule}
\omit&height6pt&\omit&&\omit&\cr %Extra space
&& $4 $
&& $\left(1-2^{-s}+2^{1-2s}\right)\left(1-2^{-s}-2^{1-3s}+2^{2-4s}\right)L_1^2(s)+L_8^2(s) $ &\cr
\omit&height6pt&\omit&&\omit&\cr %Extra space
\noalign{\hrule}
\omit&height6pt&\omit&&\omit&\cr %Extra space
&& $5 $
&& $\left(1-2^{1-s}+2^{1-2s}\right)\left(1-2.5^{-s}+5^{1-2s}\right)L_1^2(s)+\left(1+2^{1-s}+2^{1-2s}\right)L_5^2(s) $ &\cr
\omit&height6pt&\omit&&\omit&\cr %Extra space
\noalign{\hrule}
\omit&height6pt&\omit&&\omit&\cr %Extra space
&& $6$
&& $\left[1+\left(\sqrt{2}-1\right)2^{-s}+2^{1-2s}\right]\left[1-\left(\sqrt{2}+1\right)2^{-s}+2^{1-2s}\right]\left(1-2.3^{-s}+3^{1-2s}\right)L_1^2(s) +L_{12}^2(s)$ &\cr
\omit&height6pt&\omit&&\omit&\cr %Extra space
\noalign{\hrule}
\omit&height6pt&\omit&&\omit&\cr %Extra space
&& $ 7$
&& $\f{2}{3}\left(1-2^{1-s}+2^{1-2s}\right)\left(1-2.7^{-s}+7^{1-2s}\right)L_1^2(s)+\f{4}{3}\left(1+2^{-s}+2^{1-2s}\right)L_7^{-\omega}(s)L_7^{\omega^2}(s) $ &\cr
\omit&height6pt&\omit&&\omit&\cr %Extra space
\noalign{\hrule}
\omit&height6pt&\omit&&\omit&\cr %Extra space
&& $ 8$
&& $\f{1}{2}\left( 1 - 2^{1-s} + 3.2^{-2s} - 2^{2-3s} + +3.2^{1-4s} - 2^{3-5s} + 3.2^{2-6s} - 2^{4-7s} + 2^{4-8s}\right)L_1^2(s)$&\cr
\omit&height6pt&\omit&&\omit&\cr
&&
&& $+\f{1}{2}(1+2^{1-2s})L_8^2(s)+L_{16}^i(s)L_{16}^{-i}(s)$&\cr
\omit&height6pt&\omit&&\omit&\cr %Extra space
\noalign{\hrule}
\omit&height6pt&\omit&&\omit&\cr %Extra space
&& $ 9$
&& $ \f{2}{3}\left(1-2^{1-s}+2^{1-2s}\right)\left[1+\left(\sqrt{3}-1\right)3^{-s}+3^{1-2s}\right]\left[1-\left(\sqrt{3}+1\right)3^{-s}+3^{1-2s}\right]L_1^2(s)
$ &\cr
\omit&height6pt&\omit&&\omit&\cr
&&
&& $+\f{4}{3}\left(1+2^{-s}+2^{1-2s}\right)L_9^{-\omega}(s)L_9^{\omega^2}(s)$&\cr
\omit&height6pt&\omit&&\omit&\cr %Extra space
\noalign{\hrule}
\omit&height6pt&\omit&&\omit&\cr %Extra space
&& $10 $
&& $\f{1}{2}\left[1+\left(\sqrt{2}-1\right)2^{-s}+2^{1-2s}\right]\left[1-\left(\sqrt{2}+1\right)2^{-s}+2^{1-2s}\right]\left(1-2.5^{-s}+5^{1-2s}\right)L_1^2(s)
 $ &\cr
\omit&height6pt&\omit&&\omit&\cr
&&
&& $+\f{1}{2}\left[1+\left(-\sqrt{2}+1\right)2^{-s}+2^{1-2s}\right]\left[1+\left(\sqrt{2}+1\right)2^{-s}+2^{1-2s}\right]L_{5}^{2}(s)+L_{20}^i(s)L_{20}^{-i}(s)$&\cr
\omit&height6pt&\omit&&\omit&\cr %Extra space
\noalign{\hrule}
\omit&height6pt&\omit&&\omit&\cr %Extra space
&& $11 $
&& $\f{2}{5}\left(1-2^{1-s}+2^{1-2s}\right)\left(1-2.11^{-s}+11^{1-2s}\right)L_1^2(s)
$ &\cr
\omit&height6pt&\omit&&\omit&\cr
&&
&& $+\f{4}{5}\left[1+(1-\sqrt{5})2^{-1-s}+2^{1-2s}\right]L_{11}^{\tau^2}(s)L_{11}^{-\tau^3}(s)+\f{4}{5}\left[1+(1+\sqrt{5})2^{-1-s}+2^{1-2s}\right]L_{11}^{-\tau}(s)L_{11}^{\tau^4}(s)$&\cr
\omit&height6pt&\omit&&\omit&\cr %Extra space
\noalign{\hrule}
\omit&height6pt&\omit&&\omit&\cr %Extra space
&& $12 $
&& $\f{1}{2}\left(1-2,3^{-s}+3^{1-2s}\right)\left(1-2^{-s}+2^{1-2s}\right)\left(1-2^{-s}-2^{1-3s}+2^{2-4s}\right)L_1^2(s)
 $ &\cr
\omit&height6pt&\omit&&\omit&\cr
&&
&&$+\f{1}{2}\left(1+2,3^{-s}+3^{1-2s}\right)L_8^2(s)+\f{1}{2}\left(1+2^{1-2s}\right)L_{12}^2(s)+\f{1}{2}L_{24}^2(s)$ &\cr
\omit&height6pt&\omit&&\omit&\cr %Extra space
\noalign{\hrule}
\omit&height6pt&\omit&&\omit&\cr %Extra space
&& $13 $
&& $ \f{1}{3}\left(1-2^{1-s}+2^{1-2s}\right)\left(1-2.13^{-s}+13^{1-2s}\right)L_1^2(s)
+\f{1}{3}\left(1+2^{-1-s}+2^{1-2s}\right)L^2_{13}(s)$ &\cr
\omit&height6pt&\omit&&\omit&\cr
&&
&& $+\f{2}{3}\left(1-2^{-s}+2^{1-2s}\right)L_{13}^{\p^2}(s)L_{13}^{-\p^4}(s)
+\f{2}{3}\left(1+2^{-s}+2^{1-2s}\right)L_{13}^{-\p^2}(s)L_{13}^{\p^4}(s)$&\cr
\omit&height6pt&\omit&&\omit&\cr %Extra space
\noalign{\hrule}}}
}

\vfill\eject\bye